# Solitary wave solution to the generalized nonlinear Schrodinger equation for dispersive permittivity and permeability

Amarendra K.Sarma

*Department of Physics, Indian Institute of Technology Guwahati,*

*Guwahati-781039, India.*

We present a solitary wave solution of the generalized nonlinear Schrodinger equation for dispersive permittivity and permeability using a scaling transformation and coupled amplitude-phase formulation. We have considered the third-order dispersion effect (TOD) into our model and show that soliton shift may be suppressed in a negative index material by a judicious choice of the TOD and self-steepening parameter.

Veselago's landmark paper [1] on left-handed material predicted the possibility of a negative index material (NIM). He showed theoretically that if the dielectric permittivity $\varepsilon$ and the magnetic permeability $\mu$ are simultaneously negative, then one must take the index of refraction to be $n = -\sqrt{\varepsilon\mu/\varepsilon_0\mu_0}$, where $\varepsilon_0$ and $\mu_0$ are respectively the free space electric permittivity and magnetic permeability. Though naturally existing NIMs are yet to be discovered, astonishing progress have been made in artificially or man-made manufactured NIMs [2-9]. The first experimental demonstration of a Veselago medium done by Smith et al. [2]. Soon afterwards, the work by Pendry on perfect lens [3] represents the initial attempt to fill the gap between novel metamaterials and exciting applications. Since then, left-handed materials or NIMs have been studied in many different contexts ranging from second harmonic generation [10-11], propagation of electromagnetic waves [12-16] to modulation instability [17-20] and so on.

One very important difference between a NIM and an ordinary medium is that the latter has a constant permeability while the former has a frequency dependent or dispersive permeability. In order to find new features of ultra short pulse propagation resulted from the dispersive permeability, a re-examination of the pulse propagation equation is done by many researchers [17-23]. Recently, Scalora et al. [23] and then more rigorously by Wen et al. [17], have derived a generalized nonlinear Schrodinger equation (GNLSE) for dispersive permeability and permittivity suitable for few-cycle pulse propagation in NIMs. This newly derived equation has been studied so far mainly in the context of modulation instability only; however an attempt is made by Zhang and Yi [24] to find the exact solution of a generalized NLSE of Scalora type. Zhang and Yi have obtained the exact chirped soliton solution of the GNLSE using the so called variable parametric method. In this work we are finding the solitary wave solution to the GNLSE using a scaling transformation and coupled- amplitude-phase formulation [25]. We have included the effect of third-order dispersion (TOD) also in our studies. It is worth mentioning that a generalized NLSE including the effect of TOD and self-steepening, in the context of femtosecond solitary waves in ordinary optical fibers, was first studied by Christodoulides and Joseph [26].

Considering the diffraction effects to be negligible, and including the third-order dispersion, the generalized NLSE due to Wen et al. [17] could be put in the following form:

$$\frac{\partial A}{\partial z} = -i\frac{\beta_2}{2}\frac{\partial^2 A}{\partial t^2} + \frac{1}{6}\beta_3\frac{\partial^3 A}{\partial t^3} + i\gamma|A|^2 A + \Lambda\frac{\partial}{\partial t}\left(|A|^2\right)A \qquad (1)$$

where the group velocity dispersion(GVD), $\beta_2$, the third order dispersion(TOD), $\beta_3$, the self-phase modulation (SPM) coefficient, $\gamma$, and the self-steepening parameter, $\Lambda$, is defined, respectively, as

$$\beta_2 = \frac{1}{c\omega_0 n(\omega_0)}\left[\left(1 + \frac{3\omega_{pe}^2\omega_{pm}^2}{\omega_0^4}\right) - \frac{1}{n^2(\omega_0)}\left(1 - \frac{\omega_{pe}^2\omega_{pm}^2}{\omega_0^4}\right)^2\right] \qquad (2)$$

$$\beta_3 = -\frac{12}{n(\omega_0)}\frac{\omega_{pe}^2\omega_{pm}^2}{c\omega_0^6} \qquad (3)$$

$$\gamma = \frac{\chi^{(3)}\omega_0}{2n(\omega_0)c}\left(1 - \frac{\omega_{pm}^2}{\omega_0^2}\right) \qquad (4)$$

$$\Lambda = -\frac{\gamma}{\omega_0}\left[1 + \frac{\omega_{pe}^2\omega_{pm}^2 - \omega_0^4}{n^2(\omega_0)\omega_0^4} - \frac{\omega_0^2 + \omega_{pm}^2}{\omega_{pm}^2 - \omega_0^2}\right] \qquad (5)$$

All the parameters are evaluated at the frequency $\omega = \omega_0$. $\chi^{(3)}$ is the so called third-order susceptibility. As in Ref. [17], in this work also we have used the so-called lossless Drude model to describe the frequency dispersion of $\varepsilon$ and $\mu$, i.e.

$$\varepsilon(\omega) = \varepsilon_0 \left[1 - \frac{\omega_{pe}^2}{\omega^2}\right], \ \mu(\omega) = \mu_0 \left[1 - \frac{\omega_{pm}^2}{\omega^2}\right] \tag{6}$$

where $\omega$ is the angular frequency, $\omega_{pe}$ and $\omega_{pm}$ are the respective electric and magnetic plasma frequencies. In Fig.1 (a) we plot the variation of $n, \gamma$ and $\Lambda$ with the normalized frequency, $\omega_0 / \omega_{pe}$, for $\omega_{pm} / \omega_{pe} = 0.8$, while Fig.1 (b) depicts the corresponding variation for $\beta_2$ and $\beta_3$.

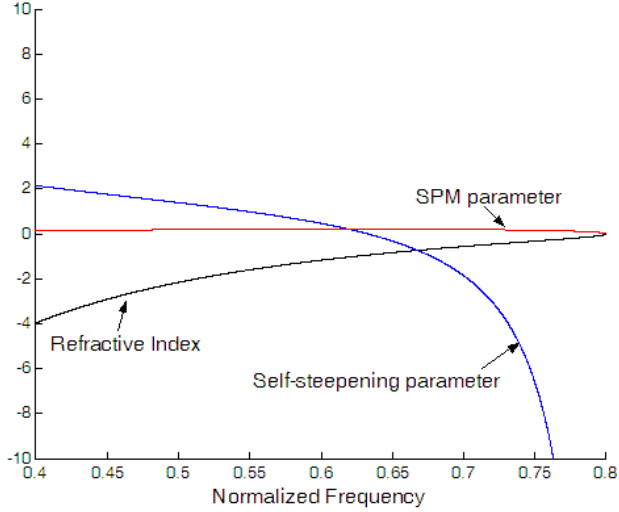

FIG.1 (a) (Color online) Variation of $n, \gamma$ and $\Lambda$ with normalized frequency for $\omega_{pe} / \omega_{pm} = 0.8$. $\gamma$ is calculated in the units of $\chi^{(3)} \omega_{pe} / c$ and $\Lambda$ is calculated in the units of $1/\omega_{pe}$.

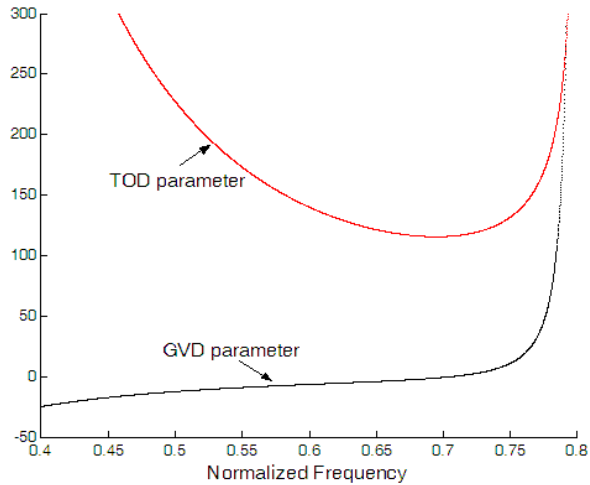

FIG.1 (b) (Color online) Variation of $\beta_2$ and $\beta_3$ with normalized frequency for $\omega_{pe} / \omega_{pm} = 0.8$, in the units of $1/c\omega_{pe}$ and $1/c\omega_{pe}^2$ respectively.

It can be easily seen that these plots are consistent with that of Ref. [17]. The TOD parameter is positive in the entire given frequency regime. In order to find the solitary wave solution of Eq. (1), we start with scaling Eq. (1), following Ref. [25], by taking: $A = b_1 u$, $z = b_2 \xi$, $t = b_3 \tau$, where $b_1, b_2$ and $b_3$ are so chosen that the coefficients corresponding to GVD, TOD and SPM become unity. Eq. (1) can now be rewritten in the anomalous dispersion regime as

$$\frac{\partial u}{\partial \xi} = i \frac{\partial^2 u}{\partial \tau^2} + i|u|^2 u + \frac{\partial^3 u}{\partial \tau^3} + \Gamma \frac{\partial}{\partial \tau}\left(|u|^2 u\right) \tag{7}$$

with $\Gamma = \dfrac{3\Lambda|\beta_2|}{\gamma \beta_3}$ \hfill (8)

We assume a solution to the Eq. (7) in the form

$$u(\xi,\tau) = U(\tau + \alpha\xi)\exp\left[i(\kappa\xi - \Omega\tau)\right] \tag{9}$$

where $U(\eta)$, with $\eta = \tau + \alpha\xi$, is a real quantity. $\alpha$ is a real parameter to be determined later. Substituting (9) in (7) and then separating the real and imaginary parts, we obtain

$$(1 - 3\Omega)U_{\eta\eta} = (\kappa + \Omega^2 - \Omega^3)U + (1 - \Omega\Gamma)U^3 \tag{10}$$

$$U_{\eta\eta\eta} = (\alpha + 3\Omega^2 - 2\Omega)U_\eta - 3\Gamma U^2 U_\eta \tag{11}$$

where $U_\eta$ is the first derivative of $U$ with respect to $\eta$ and so on.

Eqs. (10) and (11) are consistent only if the following conditions are satisfied

$$(\alpha + 3\Omega^2 - 2\Omega) = \frac{\kappa + \Omega^2 - \Omega^3}{1 - 3\Omega} \tag{12}$$

$$\Omega = \frac{1 + \Gamma}{4\Gamma} \tag{13}$$

From Eqs. (12) and (13) we obtain

$$\kappa = (\alpha + 3\Omega^2 - 2\Omega)(1 - 3\Omega) - \Omega^2 + \Omega^3 \tag{14}$$

$$\Omega = \frac{1}{4} + \frac{\gamma\beta_3}{12\Lambda|\beta_2|} \tag{15}$$

One can easily obtain the following ordinary differential equation from Eq. (11)

$$U_{\eta\eta} - \sigma U + \Gamma U^3 = 0 \tag{16}$$

with $\sigma = \alpha + 3\Omega^2 - 2\Omega$.

Eq. (16) can be solved very easily by using standard methods to obtain the following shape preserving solution:

$$u(\xi,\tau) = \sqrt{\frac{2\sigma}{\Gamma}} \operatorname{sech}\left[\sqrt{\sigma}(\tau + \alpha\xi)\right] \exp\left[i(\kappa\xi - \Omega\tau)\right] \tag{17}$$

It is important to note from Eq. (17) that localized bright soliton solutions exist only if $\sigma > 0$ and $\Gamma > 0$. In view of these constraints our analysis shows that bright soliton solution does not exist for $\omega_0/\omega_{pe} > 0.625$ with $\omega_{pe}/\omega_{pm} = 0.8$. From Eq. (17) the soliton peak power, $P_0$ and the soliton width $T_0$ can easily be estimated, after some simple algebra, as follows:

$$P_0 = \frac{9|\beta_2|^3 \sigma}{\gamma \beta_3^2 \Gamma \chi^{(3)}} \tag{18}$$

$$T_0 = \frac{\beta_3}{3|\beta_2|} \frac{1}{\sqrt{\alpha + 3\Omega^2 - 2\Omega}} \tag{19}$$

For a given $T_0, \beta_2$ and $\beta_3$ one can easily work out the parameter $\alpha$ from Eq. (19).

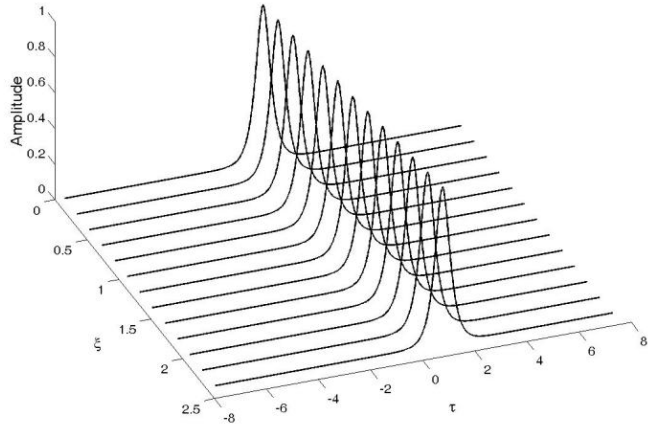

FIG.2 Spatio-temporal evolution of a soliton pulse in a NIM based waveguide with $\omega_0 = 0.6\omega_{pe}, \lambda_{pe} = 1\,\mu m$ and $T_0 = 100$ fs

In Fig.2 we depict the spatio-temporal evolution of a soliton inside the waveguide over a distance of nearly 6.5 $\mu m$ for the following typical parameters:

$\omega_0 = 0.6\omega_{pe}, \lambda_{pe} = 1\,\mu m, T_0 = 100$ fs.

It can be observed that the soliton is shifted towards the positive time axis during its evolution owing to a slight positive self-steepening (SS) effect (~0.46 in normalized unit at $\omega_0 = 0.6\omega_{pe}$). It would be interesting to check two extreme cases carefully. In Fig.3 and 4 we plot the input and output amplitude profiles of soliton pulses for $\omega_0 = 0.62\omega_{pe}$ and $\omega_0 = 0.01\omega_{pe}$ respectively. The distance of propagation in the first case is 6.5 $\mu m$ while in the later it is 20 nm.

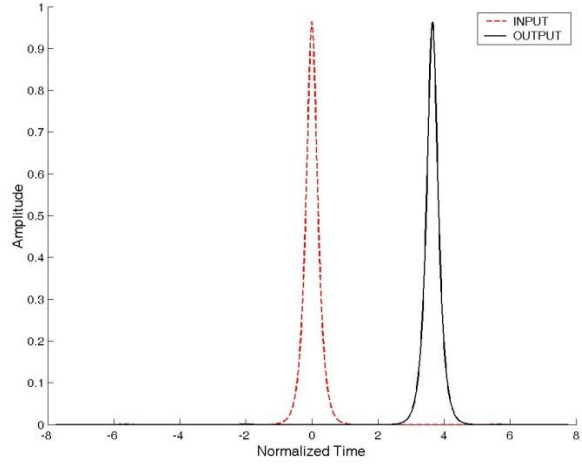

FIG.3 (Color online) Input and output soliton pulse in a NIM based waveguide with $\omega_0 = 0.62\omega_{pe}, \lambda_{pe} = 1\mu m$ and $T_0 = 100$ $fs$

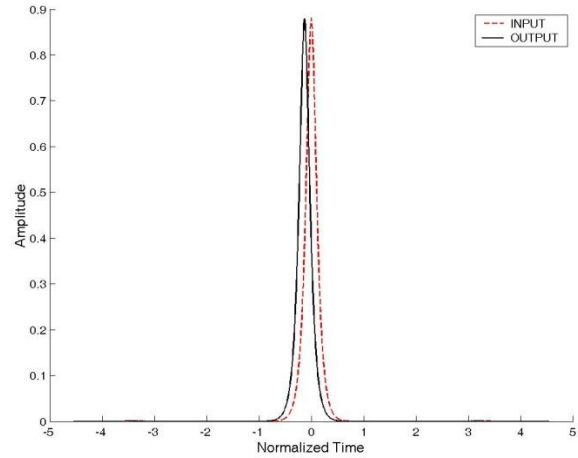

FIG.4 (Color online) Input and output soliton pulse in a NIM based waveguide with $\omega_0 = 0.01\omega_{pe}, \lambda_{pe} = 1\mu m$ and $T_0 = 100$ $fs$

It has been predicted in Ref. [19] with regard to the negative index material and other literatures [27] in the context of ordinary material, that as the propagation distance increases, the positive SS shifts the center of the generated pulse towards the trailing edge. The higher the value of SS the larger would be the shift. We observe that even though SS~ 0.19 in Fig.3 as against 0.46 in Fig.2, the shift of the soliton pulse is larger for the same propagation distance of the waveguide. It is important to note that the respective values of the TOD coefficient are 140 and 131 (in normalized units) for $\omega_0 = 0.6\omega_{pe}$ and $\omega_0 = 0.62\omega_{pe}$. Again in Fig.4 we find that the soliton is shifted towards the leading edge in spite of having a large positive SS (~100), but even larger TOD coefficient (~$10^9$) for $\omega_0 = 0.01\omega_{pe}$. From these results it may not be unreasonable to conclude that positive SS has the tendency to shift the center of the pulse towards the trailing edge while positive TOD does the opposite. In an ordinary waveguide like optical fiber, positive TOD slows down the soliton during its propagation and thereby shifts it towards the trailing edge.

In conclusion, we have found the solitary wave solution of the generalized Nonlinear Schrodinger equation including the third-order dispersion term, using a scaling transformation and coupled amplitude-phase formulation. Our study shows that the third order dispersion shifts the center of the soliton towards the leading edge of the pulse unlike in an ordinary waveguide. Hence a judicious choice of the TOD and SS parameter might be effective in controlling the soliton shift in a NIM based waveguide. This work may motivate the researchers to have a closer look at the role of the higher order dispersive effects in nonlinear pulse propagation and other related issues in NIMs.